\documentclass[prl,aps,twocolumn,showpacs,superscriptaddress,%
floatfix]{revtex4}
\usepackage{graphicx}
\usepackage{bm}
\usepackage{amsmath}
\begin{document}
\title{Half-skyrmion picture of single hole doped high-T$_c$ cuprate}
\author{Takao Morinari}
\affiliation{Yukawa Institute for Theoretical Physics, Kyoto
University, Kyoto 606-8502, Japan}
\date{\today}
\begin{abstract}
The single hole doped CuO$_2$ plane
is studied by the O(3) non-linear $\sigma$ model and its CP$^1$
representation.
It is argued that the spin configuration
around a Zhang-Rice singlet is described by a half-skyrmion.
The form of the excitation spectrum of the half-skyrmion is the same
as that in the $\pi$-flux phase.
\end{abstract}
\maketitle

In the phase diagram of high-temperature superconductors,
the most established phase is apparently the antiferromagnetic
long-range ordered phase, or the N{\' e}el ordered phase
 of undoped parent compounds.
Due to a large charge-transfer gap\cite{IFT}, a hole at each 
copper site in the CuO$_2$ plane is localized to form a spin $S=1/2$
moment.
There is an antiferromagnetic superexchange interaction $J$
between the copper site spins and the system is described by the
antiferromagnetic Heisenberg model on the square lattice,
$H_{\rm AFH}=J\sum_{\langle i,j\rangle} {\bf S}_i \cdot {\bf S}_j$.
Antiferromagnetic spin wave dispersions observed by neutron scattering
experiments are in
quite good agreement with the spin wave theory with the value of $J$
determined by Raman scattering\cite{MASON}.
High-temperature superconductivity occurs upon moderate hole doping 
on the N{\' e}el ordering phase.

For understanding the mechanism of high-temperature
superconductivity, 
it is necessary to describe the doped holes.
However, for the purpose of figuring out the proper picture,
considering a moderately doped system does not appear
to be promising because such a system is quite complicated and 
indeed there is no consensus on
the description of the system.
Contrastingly, a system with only one doped hole is much simpler
and there seems to be much hope to understand the physics
because it is closest to the well-established antiferromagnetic
long-range ordered phase.
Experimentally angle resolved photoemission spectroscopy (ARPES)
\cite{WELLS_ETAL,RONNING_ETAL,SHEN_RMP} on undoped
compounds provide valuable information on the single hole doped
system\cite{SHEN_RMP}.

As an effective theory of the antiferromagnetic Heisenberg model
$H_{\rm AFH}$, the O(3) non-linear $\sigma$ model (NL$\sigma$M) has
been studied extensively\cite{CHN}.
The model is derived from $H_{\rm AFH}$ by applying Haldane's
mapping\cite{HALDANE}:
\begin{equation}
S=\frac{\rho_s^0}{2} \int_0^{\beta} d\tau \int d^2 {\bf r}
\left[ \left( \nabla {\bf n} \right)^2
+ \frac{1}{c_{\rm sw}^2} 
\left( \frac{\partial {\bf n}}{\partial \tau} \right)^2
\right],
\label{eq_nlsm}
\end{equation}
where $\rho_s^0$ is the bare spin stiffness constant
on the length scale of the lattice constant,
which is taken to be unity,
and $c_{\rm sw}$ is the spin-wave velocity.
The integral with respect to the imaginary time $\tau$ 
is carried out up to the inverse temperature $\beta$.
(Hereafter we take the unit $\hbar = 1$.)
The unit vector ${\bf n}$ represents the directions of staggered spin
moments.
The low-energy and the long-length scale behaviors of $H_{\rm AFH}$
are well described by NL$\sigma$M as confirmed 
by experiments \cite{MASON} and quantum Monte Calro
simulations\cite{QMC}.

There is a useful form for the NL$\sigma$M, which is called CP$^1$
model\cite{RAJARAMAN}. 
Introducing complex fields $\zeta_{\sigma}(x)$ with $x=(c_{\rm
sw}\tau, {\bf r})$ through 
${\bf n}=\sum_{\sigma,\sigma=\uparrow,\downarrow} 
\overline{\zeta}_{\sigma} 
{\mbox{\boldmath ${\bf \sigma}$}}_{\sigma\sigma'}
\zeta_{\sigma'}$
with $\sigma^{\alpha}$($\alpha=x,y,z$) representing Pauli matrices,
we obtain $S=(\rho_s^0/2c_{\rm sw})\int d^3 x \left[ |\partial_{\mu}
\zeta_{\sigma}|^2+\left(\overline{\zeta}_{\sigma}
\partial_{\mu} \zeta_{\sigma} \right)^2 \right]$.
Performing a Stratonovich-Hubbard transformation for the second term
in the square brackets, we obtain the CP$^1$ model:
\begin{equation}
S=\frac{1}{g}\int d^3 x |\left( \partial_{\mu} -i \alpha_{\mu}
\right)\zeta_{\sigma}(x)|^2,
\label{eq_cp1}
\end{equation}
where $g=c_{\rm sw}/2\rho_s^0$ and 
$\langle \alpha_{\mu}(x) \rangle=  
\langle i\sum_{\sigma} \overline{\zeta}_{\sigma}(x)\partial_{\mu}
\zeta_{\sigma} (x) \rangle$
at the saddle point.
The field $\alpha_{\mu}$ is a U(1) gauge field associated with a
local gauge symmetry, 
$\zeta_{\sigma}(x) \rightarrow \zeta_{\sigma}(x) \exp(i\chi(x))$.
Thus, the NL$\sigma$M is equivalent to the boson system with the U(1)
gauge field interaction.

The CP$^1$ model is also derived from 
the Schwinger boson mean field theory (SBMFT)\cite{AA}.
In the theory, the spin is represented in terms of boson operators
$z_{j\sigma}^{\dagger}$ and $z_{j\sigma}$ as 
${\bf S}_j =\frac12 \sum_{\sigma,\sigma'}
z_{j\sigma}^{\dagger}
{\mbox{\boldmath ${\bf \sigma}$}}_{\sigma \sigma'}
z_{j\sigma'}$
with the constraint $\sum_{\sigma=\uparrow,\downarrow}
z_{j\sigma}^{\dagger} z_{j\sigma}=1$.
The SBMFT Hamiltonian is given by 
$H_{\rm SB} = -(J/2)\sum_{\langle i,j \rangle}
\left[ \Delta^* \left( z_{i\uparrow}z_{j\downarrow} 
- z_{i\downarrow}z_{j\uparrow} \right) + h.c. -|\Delta|^2
\right]+\sum_{j\sigma} \lambda 
\left(z_{j\sigma}^{\dagger} z_{j\sigma}-1\right)$,
with $\Delta$ and $\lambda$ being the mean field parameters.
Here $\Delta = \langle z_{i\uparrow}z_{j\downarrow} 
- z_{i\downarrow}z_{j\uparrow} \rangle$ 
describes the Schwinger boson 
pairing\cite{READ_SACHDEV,CHUBUKOV,NG95}.
After Fourier transforming,
the Schwinger boson quasi-particle excitation spectrum is obtained by
a Bogoliubov transformation\cite{YOSHIOKA}:
$\omega_k = \sqrt{\lambda^2 - J^2|\Delta|^2 
\left( \sin k_x \pm \sin k_y \right)^2}$,
where the plus sign is for $k_x k_y >0$ and
the minus sign is for $k_x k_y <0$.
In the N{\' e}el ordering phase, $\lambda=2J|\Delta|$ and
the Schwinger bosons are gapless at zone face centers 
$(\pm \pi/2,\pm \pi/2)$.
The N{\' e}el ordering state is a consequence of Bose-Einstein
condensation 
of the Schwinger bosons at those points\cite{YOSHIOKA,BEC}.
One can derive the CP$^1$ model after some
algebra\cite{READ_SACHDEV}.
The relation between the Schwinger boson fields and
$\overline{\zeta}_{\sigma}(x)$ and $\zeta_{\sigma}(x)$ 
in Eq.(\ref{eq_cp1}) is
$z_{j\sigma}=\zeta_{j\sigma}$ at one sublattice and
$\zeta_{j\sigma}=z_{j\sigma}^{\dagger}$ at the other sublattice.
One Schwinger boson carries spin $1/2$ and the unit gauge charge.
Therefore, pairs of the Schwinger bosons that describe the low-lying
excitations carry integer spins and the gauge charge two.

Both of NL$\sigma$M and SBMFT are, so to speak, bosonic description of 
the system.
Meanwhile, one can construct a fermionic theory by representing
the $S=1/2$ spins in terms of fermions.
Mean field analysis was carried out by Affleck and
Marston\cite{AFFLECK_MARSTON} and the $\pi$-flux phase was proposed.
The state is characterized by a flux $\pi$ penetrating each plaquette
with alternating directions.
The quasiparticle excitations are gapless at $(\pm \pi/2, \pm \pi/2)$.
More advanced analysis based on an effective theory that
includes fluctuations around the mean field state through a U(1) gauge 
field shows that a mass term is induced in the spectrum dynamically.
This phenomenon is associated with the N{\' e}el
ordering\cite{MARSTON,KIM_LEE}.

It has been shown experimentally that doped holes reside primarily 
on oxygen sites.
Zhang and Rice suggested\cite{ZHANG_RICE} that there is 
strong correlations of forming a singlet pair between oxygen
p-orbital holes and copper d-orbital holes.
The t-J model was proposed based on this picture.
Although the single hole problem has been studied mostly by this
model, the focus is mainly on frustration effects induced by hopping
of the Zhang-Rice singlet\cite{SHRAIMANN_SIGGIA,KANE_LEE_READ}.
However, not so much attention has been paid on the spin configuration 
around the Zhang-Rice singlet.

\begin{figure}[bp]
 \includegraphics[scale=0.8]{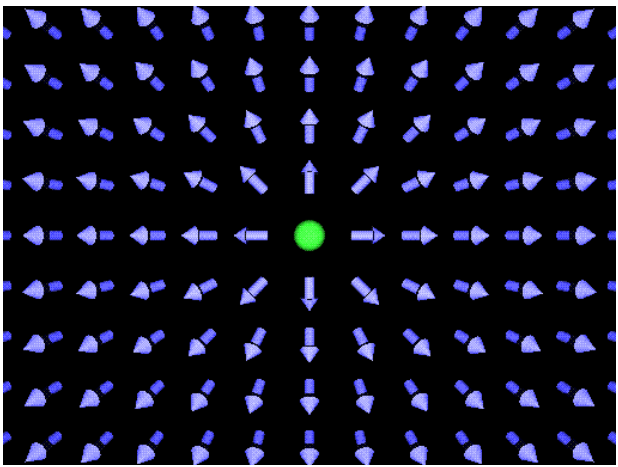}
 \caption{The half-skyrmion spin texture. Arrows indicate the
staggered moment and the filled circle at the center represents the
Zhang-Rice singlet formed site.}
 \label{fig_hs}
 \end{figure}

In this paper, it is argued that in the single hole doped system the
doped hole induces a half-skyrmion which is schematically shown in
Fig.~\ref{fig_hs}.
The half-skyrmion is a spin texture
characterized by a half of a topological charge.

We start with considering the spin configuration around a static
Zhang-Rice singlet.
The moving case shall be considered later.
A Zhang-Rice singlet is composed of a d-orbital hole state and the
four oxygen hole states around the copper ion.
Constructing the Wannier functions, the operator for the
Zhang-Rice singlet is given by\cite{ZHANG_RICE}
$\frac{1}{\sqrt{2}} 
\left( \phi_{j\uparrow} d_{j\downarrow} - \phi_{j\downarrow}
d_{j\uparrow} \right)$,
where $\phi_{j\sigma}$ is the symmetric
combination state of the four oxygen hole states around the copper ion
at the site $j$\cite{ZHANG_RICE}.
The d-orbital state of the Zhang-Rice singlet is given by
the superposition of the up-spin state and the down-spin state.
To be specific, we assume that 
in the N{\' e}el ordered state before hole doping the ordered spin
moment at the site $j$ is in the direction of the positive $z$-axis.
Then, the d-orbital state with up-spin does not affect
the neighboring spins so much.

In contrast, the down-spin state affects the neighboring spins to
change their directions.
From the analysis of the NL$\sigma$M below, we show that a spin
configuration, which is characterized by a non-zero topological
charge, is created around the down-spin state.
This spin configuration is called skyrmion\cite{RAJARAMAN}.
For the static case, the energy is given by 
$
E=\frac{\rho_s^0}{2}\int d^2 {\bf r} 
\left( \nabla {\bf n} \right)^2.
$
The field equation is $\nabla^2 {\bf n}-\left( {\bf n} \cdot \nabla^2
{\bf n} \right) {\bf n}=0$.
Solutions are devided into sectors characterized by a topological
charge : $
Q=\frac{1}{8\pi} \int d^2 {\bf r} \epsilon_{\mu \nu} {\bf n} \cdot
\left( \partial_{\mu} {\bf n} \times \partial_{\nu} {\bf n} \right)$,
where $\epsilon_{xy}=-\epsilon_{yx}=1$ and
$\epsilon_{xx}=\epsilon_{yy}=0$.
The energy in each sector has the lower bound:
$E\geq 4\pi \rho_s^0 |Q|$.
The equality is satisfied if and only if
$\partial_{\mu} {\bf n} = \pm \epsilon_{\mu \nu} \left( {\bf n} \times 
\partial_{\nu} {\bf n} \right)$\cite{BELAVIN_POLYAKOV}.
The lowest enegy state is obtained by solving this equation with the
boundary conditions: 
${\bf n}\rightarrow +{\hat e}_z$ at infinity and 
${\bf n}= -{\hat e}_z$ at ${\bf r} = {\bf r}_j$.
In terms of a variable $w\equiv (n_x + i n_y)/(1-n_z)$, the equation 
turns out to be the Chaucy-Riemann
equations\cite{BELAVIN_POLYAKOV}.
Therefore, $w$ is an analytic function of $z=x+iy$ or 
$z^*=x-iy$.
The lowest energy state satisfying the boundary conditions is
$w_s=[(x-x_j)+i(y-y_j)]/\lambda$ and
$w_{as}=[(x-x_j)-i(y-y_j)]/\lambda$ with
$\lambda$ being a constant.
In the vector ${\bf n}$ representation, these solutions are
\begin{equation}
{\bf n}=\left(
\frac{2\lambda (x-x_j)}{|{\bf r}-{\bf r}_j|^2+\lambda^2},
\pm \frac{2\lambda (y-y_j)}{|{\bf r}-{\bf r}_j|^2+\lambda^2},
\frac{|{\bf r}-{\bf r}_j|^2-\lambda^2}{|{\bf r}-{\bf
r}_j|^2+\lambda^2}
\right),
\label{eq_sky}
\end{equation}
where the plus sign is for $\omega_s$ and the minus sign is for
$w_{as}$.

Now we consider the superposition of the uniform state 
${\bf n}=+{\hat e}_z$ for the up-spin state and the
skyrmion spin configuration (\ref{eq_sky})
created by the down-spin state.
The NL$\sigma$M analysis suggests that a spin
configuration with non-zero topological charge $Q$, where $0<|Q|<1$,
is formed with 
$n_{\ell z}\geq 0$ for any site $\ell$.
The value of the topological charge $Q$ is determined 
through the flux representation in terms of the 
CP$^1$ gauge field $\alpha_{\mu}$:
$
Q=\frac{1}{2\pi} \int d^2 {\bf r} \left( \nabla \times 
{\mbox{\boldmath ${\bf \alpha}$}} \right).
$
The spin configuration with $Q$
corresponds to the flux $2\pi Q$.
Due to Bose-Einstein condensation of the bosons $\zeta_{\sigma}(x)$,
the value of the flux $2\pi Q$ is not arbitrary.
Since low-lying excitations are pairs of these bosons, the flux
quantum is $\pi$ analogous to conventional BCS superconductors.
From the constraint on $Q$, i.e., $0<|Q|<1$, we conclude that $2\pi
|Q|=\pi$, that is, $|Q|=1/2$.
Therefore, the spin configuration is 
the half-skyrmion, which is equivalent to the skyrmion with the
core, and its energy is $E_0=2\pi\rho_s^0$.
This conclusion can be reached if we assume that the staggered
magnetization vanishes at the perimeter of the core. 
The analysis of the NL$\sigma$M gives the half-skyrmion solution
as was shown by Saxena and Dandoloff\cite{SAXENA_DANDOLOFF}.
They considered the two-dimensional ferromagnetic Heisenberg model
in the context of the quantum Hall systems.
For the static case, the field equation is the same.
So we can apply their analysis to the antiferromagnetic case as well.


The moving half-skyrmion is
obtained from the static half-skyrmion by applying 
a Lorentz transformation
because the action (\ref{eq_nlsm}) and (\ref{eq_cp1}) are Lorentz
invariant with $c_{\rm sw}$ the speed of ``light.''
From the calculation of the energy-momentum tensors, we find
$\epsilon_k^0 = \sqrt{c_{\rm sw}^2 (k_x^2+k_y^2) 
+ E_0^2}$.
For the $S=1/2$ antiferromagnet, the bare spin-wave velocity 
is $c_{\rm sw}=\sqrt{2}J$ and the bare spin stiffness is 
$\rho_s^0=J/4$.
Renormalization effect leads to 
$c_{\rm sw} \rightarrow \sqrt{2}Z_c J$
and $\rho_s^0 \rightarrow Z_{\rho} J/4 (\equiv \rho_s)$\cite{CHN}.
Thus, the renormalized dispersion is given by
$\epsilon_k = J\sqrt{2Z_c^2 (k_x^2+k_y^2) + (\pi Z_{\rho}/2)^2}$.
Since the doped hole is fermion, 
this relativistic dynamics of the half-skyrmion is described by
a Dirac fermion Lagrangian density:
\begin{equation}
{\cal L} =\sum_{\sigma} \overline{\psi}_{\sigma}
\left( \gamma_{\mu} \partial_{\mu} + m c_{\rm sw}^2\right)
\psi_{\sigma},
\label{eq_dirac}
\end{equation}
with $\overline{\psi} = \psi^{\dagger} \gamma_0$ and 
$mc_{\rm sw}^2=\pi Z_{\rho}J/2$.
The $\gamma$ matrices satisfy 
$\left\{ \gamma_{\mu}, \gamma_{\nu} \right\}=2 \delta_{\mu \nu}$
in the Euclidean space time.
Here $\sigma$ is an index for the sign of the topological charge.
The number of components of the Dirac fermion field is 
four because there are a positive energy state and negative energy
state and the origin of the Dirac fermion dispersion is
at either ${\bf k}_1=(\pi/2, \pi/2)$ or 
${\bf k}_2=(-\pi/2,\pi/2)$.
The latter is implied from the fact that the Schwinger bosons are
gapless at those points in the symmetry broken N{\' e}el ordering
phase.
Note that the dynamics is associated with the
properties of the parent compound, namely, the Mott insulator.
In this sense, the fermion $\psi$ is different from
quasi-particles in the conventional Fermi liquid.

The lattice version of the action is derived by discretizing
Eq.(\ref{eq_dirac}).
As is well-known in the lattice gauge field theory\cite{KOGUT83},
there is a fermion-doubling problem when one introduces a lattice 
for a Dirac fermion.
This subtlity does not make any difficulty in our case because of the
antiferromagnetic correlations that naturally introduce two species
of Dirac fermions with half of the Brillouin zone (the magnetic
Brillouin zone).
Discretizing Eq.(\ref{eq_dirac}) and then performing Fourier
transformation, we obtain
$
{\cal L}=\sum_{\sigma} \overline{\psi}_{k\sigma}
\left(
\begin{array}{cc}
mc_{\rm sw}^2+\partial_{\tau} & \cos k_x + i \cos k_y \\
-\cos k_x + i\cos k_y & mc_{\rm sw}^2-\partial_{\tau}
\end{array}
\right)
\psi_{k\sigma}.
$
Here we have used the fact that the origin of the Dirac fermion
dispersion is at ${\bf k}_{1,2}$.
The dispersion relation is given by
\begin{equation}
\epsilon_k^{\pm} = \pm \sqrt{c_{\rm sw}^2 
(\cos^2 k_x + \cos^2 k_y) + (mc_{\rm sw})^2}.
\label{eq_disp}
\end{equation}
The dispersion curve $\epsilon_k^- - mc_{\rm sw}^2$
is qualitatively in good agreement with the experimentally
observed dispersion by Wells {\it et al}\cite{WELLS_ETAL,LAUGHLIN97}.
In our theory, $Z_c$ and $Z_{\rho}$ are determined by the parameters
of the undoped system if one neglects spin wave effects on the
self-energy of the half-skyrmion.
For the $S=1/2$ antiferromagnetic Heisenberg model,
the values of $Z_c$ and $Z_{\rho}$ are estimated as
$Z_c = 1.17$ and $Z_{\rho} = 0.72$
from quantum Monte Calro simulations\cite{QMC} 
and a series expansion analysis\cite{SINGH}.
Substituting these values into Eq.~(\ref{eq_disp}),
we find that the band width along $(0,0)$ to $(\pi,\pi)$
is $\sim 1.5J$ and the Dirac fermion mass is $m \sim 1.13J$.
Experimentally, the band width is estimated to
$2.2J$\cite{WELLS_ETAL}.
This $30\%$ discrepancy would be associated with the deviation of the
real system from the NL$\sigma$M description.
There is also self-energy corrections on the mass $m$
due to spin waves.
We have estimated it to $m/J = 1.13/(1-0.12 e_A^2)$
where $e_A$ is a coupling parameter between the half-skyrmion and
the antiferromagnetic spin waves.
The coupling parameter $e_A$ turns out to be a gauge charge in the
dual theory below.


The effective action for the half-skyrmion can be derived by 
duality mapping\cite{FISHER_LEE}.
In the CP$^1$ theory, the half-skyrmion solution residing at the
origin has the following form
$
z_{\sigma}(x)=\frac{\lambda u_{\sigma}+rv_{\sigma}
\exp(\pm i\theta)}{\sqrt{r^2+\lambda^2}},
$
where ${\bf u}^* \cdot {\bf v}=0$, and $|{\bf u}|=|{\bf v}|=1$.
(For the solution (\ref{eq_sky}) with setting ${\bf r}_j=0$,
these parameters are
$v_{\uparrow}=u_{\downarrow}=1$ and 
$u_{\uparrow}=v_{\downarrow}=0$.)
If we take $|v_{\uparrow}|=|v_{\downarrow}|\neq 0$, then the 
staggered moments lie in the plane with the same direction at
infinity.
In this boundary condition, $z_{\sigma}(x)\sim v_{\sigma}
\exp(\pm i\theta)$ at $r\gg \lambda$.
We write $z_{\sigma}(x)=\rho_{\sigma}^{1/2} \exp(i\phi(x))$,
where $\phi(x)=\phi_0(x) +\phi_v(x)$.
Here $\phi_0(x)$ is a non-singular function of $x$ and $\phi_v(x)$
describes the vortex (half-skyrmion) field:
$\phi_v(x) \sim q_v \tan^{-1} (y-y_v)/(x-x_v)$,
where $q_v$ is the sign of the topological charge and 
$(x_v,y_v)$ denotes the position of the vortex.
(Note that $x_v$ and $y_v$ are functions of the imaginary time.)
In applying the duality mapping, a dual gauge field is introduced.
The coupling between the gauge field and the half-skyrmion
has the form of the minimal coupling.
Thus, we obtain the following QED$_3$ Lagrangian density in the
continuum,
\begin{equation}
{\cal L} = \sum_{\sigma} \overline{\psi}_{\sigma}
\left[
\gamma_{\mu} \left( \partial_{\mu} 
- i q_{\sigma} A_{\mu} \right)
+ m \right]
\psi_{\sigma} + \frac{1}{4e_A^2} 
\left( \partial_{\mu} A_{\nu} - \partial_{\nu} A_{\mu}
\right)^2.
\label{eq_qed3}
\end{equation}
The gauge field $A_{\mu}$
originally comes from the spin current, which is
associated with the antiferromagnetic spin wave modes.
The gauge charge $e_A$ is a parameter of the theory.
Estimation of $e_A$ is beyond the analysis of the effective theories.
For analysis of the half-skyrmion self-energy,
we formulate the theory on the lattice according to a standard
procedure\cite{KOGUT83}.
The second order term with respect to 
the one ``photon'' vertex part is evaluated numerically and the first
order term with respect to the two ``photon'' vertex part is computed 
analytically.
While the former has negligible effect, the latter leads to
$\Sigma_k = -(i\sqrt{2}e_A^2/16)\sum_{\mu} \gamma_{\mu}
\sin \left( k_{\mu} a \right)
\int_{-\pi}^{\pi} \frac{d^2 \tilde{{\bf k}}}{(2\pi)^2}
\frac{1}{\sqrt{1-(\cos \tilde{k}_x + \cos \tilde{k}_y)^2/4}}$.
The mass renormalization due to this term is
$m \rightarrow m/(1-0.12 e_A^2)$.

A similar action of Eq.(\ref{eq_qed3}) 
is obtained in the fermionic
representation of the $S=1/2$ antiferromagnetic Heisenberg model 
\cite{MARSTON,KIM_LEE} based
on the $\pi$-flux phase\cite{AFFLECK_MARSTON} with dynamically induced 
mass $m$, through the interaction with a gauge field.
(The condition of this dynamical mass generation\cite{APPELQUIST_ETAL}
is that the number of Dirac fermion species
is less than $32/\pi^2 \sim 3.2$.
This is the case for the QED$_3$ theory of the
antiferromagnet\cite{KIM_LEE,TESANOVIC}.)
From exact diagonalization studies it was shown \cite{GOODING} 
that the bond spin currents which characterize the $\pi$-flux phase is
reproduced by a skyrmion-like spin texture\cite{SKYRMIONS}.
From the view point of the spin-charge separation\cite{ANDERSON},
Baskaran suggested\cite{BASKARAN} that a half-skyrmion can be seen
as deconfined spinons based on an analysis of an $n$-skyrmion
solution.

In summary, we have argued that, within the effective theory
approaches to the $S=1/2$ Heisenberg antiferromagnet, the doped hole
induces the half-skyrmion spin texture through the formation of the
Zhang-Rice singlet in the single hole doped system.
This picture is consistent with rapid suppression of the N{\' e}el
ordering by the doped holes because the half-skyrmion behaves as a
vortex in the condensate.
It would be interesting to extend the picture to the slightly doped
regime.

I would like to thank G.~Baskaran, X.~Dai, 
M.P.A.~Fisher, L.~Balents, N.~Nakai,
K.~Shizuya, and A.~Tanaka for valuable discussions. 
I also thank Z.~Te\u{s}anovi\'{c} for helpful comments on the
manuscript.
This work was supported in part by the Grant-in-Aid for the 21st
Century COE "Center for Diversity and Universality in Physics" from
the Ministry of Education, Culture, Sports, Science and Technology
(MEXT) of Japan.

\subsection{}
\subsubsection{}

\end{document}